\documentclass[oneside]{book}

\usepackage{xcolor}
\usepackage{amsmath}
\usepackage{amssymb}
\usepackage{mathtools}
\usepackage{titling}
\usepackage{ragged2e}

\usepackage{bm} 
\usepackage{booktabs}
\usepackage{caption} 
\usepackage{dcolumn}
\usepackage{fancyhdr} 
\usepackage{graphicx}
\usepackage{subfig}
\usepackage{multirow}
\usepackage[titletoc,toc,page]{appendix}

\usepackage{mathpazo} 
\usepackage{hyperref}
\usepackage{latexsym}

\usepackage[symbol]{footmisc}
\usepackage{emptypage}

\usepackage{graphicx}  % needed for figures
\graphicspath{ {./images/} }

\usepackage{dcolumn}   % needed for some tables
\usepackage{bm}        % for math
\usepackage[square,sort,comma,numbers]{natbib}

\makeatletter
\usepackage{xparse}
\usepackage{etoolbox}

\ExplSyntaxOn
\NewDocumentCommand{\definechapternumbers}{m}
 {
  \clist_gset:Nn \g_raihaneh_chapters_clist { #1 }
 }
\RenewExpandableDocumentCommand{\thechapter}{}
 {
  \clist_item:Nn \g_raihaneh_chapters_clist { \value{chapter} }
 }
\ExplSyntaxOff

\definechapternumbers{2}

\usepackage{titlesec}
\usepackage{lipsum}
\titleformat{\chapter}[display]{\normalfont\bfseries}{}{0pt}{\Large}
\fancypagestyle{plain}{%
\fancyhf{}
\fancyhead[L]{\textbf{NEW ERA SCIENCE \\ Vol. 1. Issue No. 1, June, 2025\\ ISSN 3082-5784}}
}

\begin{document}
\fancyhf{}
\fancyhead[R]{\textbf{NEW ERA SCIENCE (ISSN 3082-5784)\\ Volume 1, Issue No. 1, June, 2025\\[25pt]}}
\section*{}

\vspace{35pt}
% the following line is for submission, including submission to the arXiv!!
%\hspace{5.2in} \mbox{Fermilab-Pub-04/xxx-E}
\begin{center}\textbf{\large Towards an Entropic Geometrodynamics of Quantum Particles:\\
Mouths of Casimir Wormholes as Planckeons} \\[20pt]
Jeffrey Alloy Q. Abanto\\  Astronomy Department, New Era University
\end{center}
%\maketitle
\vspace{5pt}
\begin{center}\textbf{Abstract}\end{center}
\noindent A conformal gravity approach is presented here to describe the emergence of a quantized spacetime at the Planck scale from a lattice-like network of mouths of traversable wormholes as planckeons. This differs from the earlier work of Licata et.al., which started with the assumption that ER=EPR conjecture is valid and used an entanglement entropy from the Ryu-Takayanagi formula. Here, we developed first a conformal gravity model that is postulated to apply at the Planck scale to describe the spacetime fluctuation via a second-order form of the Ricci Flow, i.e., a wave equation of the metric tensor. We then consider its black hole solution to suggest the possible formation of Planckian wormholes. By considering the role of the entropic force due to Casimir effect as the source of negative energy, we suggest that Planckian wormholes will be stable and may allow for the transfer of field through a series of Casimir wormholes with mouths that are open at a given period of time as dictated by a modified Uncertainty Principle. From this phenomena at Planck scale, we showed that it can be the basis from which "quantum dynamics" emerges and therefore ER=EPR conjencture can be realized. Lastly, we develop a new quantum interpretation that focuses on the nature of quantum dynamics, particularly with the effect of generating the famous interference pattern in a macroscopic double-slit experiment.  We showed that the interference pattern's origin is entropic in origin and chaotic in nature, i.e., it is a phenomenon that is sensitive to initial conditions, such as both the interaction of the particle with the slit and the non-interation of it by overcoming the barrier using a wormhole. However, the dynamics of a quantum particle through a series of wormholes is not just considered to be probabilistic in nature like in a Brownian motion but also similar to the case of the Traveling Salesman Problem that requires an NP-hard combinatorial optimization.  \\[10pt]
\noindent \textbf{Keywords:} Planckeons, ER=EPR Conjencture, Quantum Interpretation\\[10pt]

%\title{On Geometrodynamics of Quantum Particles\\\;\;\;}
%\author{Jeffrey A. Q. Abanto}
%\address{Department of Atmospheric and Space Sciences\\ Cosmology and Astrophysics Unit, New Era University\\ No.9 Central Ave.,Quezon City, 1107 Philippines\\}
%\begin{abstract}
%\noindent Quantum Mechanics will be derived here in a different way. It will be derived from a new conformal gravity theory that is postulated to apply at Planck scale and uses Wheeler's notion of a quantum foam in the description of spacetime at Planck scale. The new gravity theory will also be used as the basis of a new geometric formulation and interpretation of Quantum Mechanics where we consider the role of the entropic force and the possibility that a quantum particle, at the fundamental level, travels through a series of Planckian wormholes. Also, the new quantum interpretation proves \lq{}\lq{}ER=EPR" conjencture and suggests an entropic origin of the interference pattern in the double-slit experiment. 
%\\
%\end{abstract}
%\begin{keyword}
%Quantum Gravity, Quantum Formalism and Intepretation\\
%PACS number: 04.50.Kd,04.60.-m,03.65-w
%%\end{keyword}

\fancyfoot[C]{\thepage}
\setcounter{page}{17}
%\maketitle
%\tableofcontents

\vspace{25pt}

\section*{Introduction}
In the recent work of Licata et.al.,\cite{lic},  the mouths of a vast network of non-traversable Einstein-Rosen bridges at the Planck scale are considered as the planckeons, the Planck scale excitations endowed with minimal length and time uncertainties. By treating the lattice of planckeons as a statistical ensemble, they derived in their paper a partition function whose high temperature limit reproduces a logarithmic Bekenstein type entropy. Then, by embedding the minimal length directly into the wormhole throat, leads to a quantum corrected Bekenstein entropy that ties the information content of spacetime to the non-local correlations carried by the planckeon lattice. The results of their work, such as modified Uncertainty Principle and the wormhole metric, according to them, suggest that spacetime is an entanglement-driven condensate whose macroscopic connectivity, thermodynamics and quantum information content can all be traced back to the dynamics of a Planck scale network of wormhole mouths. In this study, we aim to use a different approach, a conformal gravity approach, that will yield almost the same results with a similar viewpoint on the nature of Wheeler's quantum foam at the Planck scale. However, in developing our modified Uncertainty Principle, we use our previous work on Deformed Special Relativity\cite{Aba2}. For the wormhole metric, we use the Shu-Shen solution\cite{ShuShe} by considering the spacetime fluctuation to be a metric fluctuation represented by a normalized wave equation of the metric tensor. Such equation is known as the Hyperbolic Geometric Flow\cite{DeKe}, which is a second-order version of the famous Ricci Flow\cite{thurs}.  The paper is organized as follows: we review first Conformal Gravity, Deformed Special Relativity and Ricci Flow. Then we introduced Hyperbolic Geometric Flow, derived all the related quantum equations and discussed the Shu-Shen solution. Lastly, we discussed an outline of a new quantum interpretation, the Topo-Metrodynamic Interpretation(TMI), that focuses on how the double-slit interference pattern could be generated by an entropic and chaotic processes via a combinatorial optimization.

\section*{Conformal Gravity Approach}
 Conformal Gravity is a term given to any alternative gravity theory that is invariant under a conformal transformation known as the Weyl transformation. Such alternative gravity theories are also called as \lq{}\lq{}Weyl-squared" theories as their Lagrangian involves square of the Weyl Tensor. The first conformal gravity theory was proposed by Weyl in the 1920s in his attempt to unify General Relativity to Maxwell's electromagnetic field theory. He considered that the scale parameter $\omega$ in a conformal transformation, $g_{\mu\nu}\rightarrow e^{2\omega}g_{\mu\nu}$, is related to electromagnetism via the gauge transformation involving the 4-vector potential, i.e., $A^{\mu}\rightarrow A^{\mu}+\partial_{x}\omega$. Weyl's theory failed, but it spurred a considerable amount of research in gravitational theories based on conformal invariance. In the late 1920s, Weyl, London and others successfully resurrected Weyl's idea when they applied it to Quantum Mechanics although the gauge transformation is no longer with the metric tensor but with the wave function. A comprehensive history and review of Weyl Geometry and its application on various topics such as Quantum Gravity and Foundations of Quantum Mechanics can be found in the work of Scholz\cite{Sch2011} and references therein. Also of great interest are the works of J.T. Wheeler \cite{JWhe}, Wood and Papini \cite{WoPa}, Mannheim\cite{mann}, Castro\cite{wg2}, Shojai et.al.\cite{wg3}, Faria\cite{faria} and the work of Maldacena\cite{mal} on Conformal Field Theory. All are very much rooted in the work initiated by Weyl in 1918.\\
\indent When Weyl evaluated the changes to Riemannian Geometry of his conformal transformation, he found out that the affine connection is modified where a pseudo-vector must be introduced. In order for the affine connection to be invariant, the pseudo-vector $\phi$ must undergo a gauge transformation: $\tilde{\phi}=\phi-d \log{\omega}$. The introduction of the pseudo-vector leads to the non-metricity condition where the covariant derivative of metric tensor is non-vanishing. The geometry that was later developed is now called Weyl Geometry where the corresponding Weyl curvature tensor (in 4 dimensions),
\begin{eqnarray}
W^{\lambda}_{\nu\alpha\beta}=R^{\lambda}_{\nu\alpha\beta}+\frac{1}{2}(\delta^{\lambda}_{\beta}R_{\nu\alpha}-\delta^{\lambda}_{\alpha}R_{\nu\beta}\nonumber +g_{\nu\alpha}R^{\lambda}_{\beta}
-g_{\nu\beta}R^{\lambda}_{\alpha})+\frac{1}{6}(\delta^{\lambda}_{\alpha}g_{\beta\nu}-\delta^{\lambda}_{\beta}g_{\alpha\nu})R 
\end{eqnarray}
\noindent is also invariant under Weyl transformation. It is now known to be important in describing compression and tidal deformation. From this tensor, a unique Lagrangian can be derived which is now called as the Weyl Lagrangian
\begin{eqnarray}
L_{w}=\sqrt{-g}W_{\mu\nu\alpha\beta}W^{\mu\nu\alpha\beta}=\sqrt{-g}(R_{\mu\nu\alpha\beta}R^{\mu\nu\alpha\beta}-2R_{\mu\nu}R^{\mu\nu}+\frac{1}{3}R^{2})\nonumber
\end{eqnarray}
\noindent which is also invariant under Weyl\rq{}s conformal transformation. It is useful, nowadays, in formulating an alternative gravity theory at macroscopic scale. Despite of all the successes of Weyl's original theory, the idea that it can be used in unifying gravity with Quantum Mechanics is not currently considered by many physicists.\\
\indent In this paper, we wanted to show that conformal invariance can still be used to unify gravity and Quantum Mechanics at the Planck scale. To do this, we also suggest the use of a tool in Differential Topology called the Hyperbolic Geometric Flow to describe the fluctuation of spacetime at the Planck scale, the so-called spacetime foam. Then we apply the Weyl transformation which involves a complex exponential function as the conformal factor. These serve as the basis of the underlying postulates of a new theory of gravity at Planck scale which can be stated as follows:
\begin{verse}
\noindent \bf {Postulate 1: Invariance under the Weyl Transformation, $g_{\mu\nu}\rightarrow\Omega^{2}g_{\mu\nu}$, is a fundamental symmetry in nature, where the conformal term $\Omega^{2}$, in general, is a complex function.}\\[10pt]
\noindent \bf{Postulate 2: Ricci Flow is a statistical system that can be used to describe the spacetime fluctuation and curvature at Planck Scale.}
\end{verse}

\vspace{25pt}
\noindent In Postulate 1, the demand for Weyl transformation to be a fundamental symmetry in nature, means that it should also be applicable at the quantum and Planck scale and not just at the macroscopic scale. At macroscopic scale, the particular type of conformal transformation used by Weyl was successful in adding additional degree of freedom in General Relativity as it explains compression and tidal forces. It failed in electromagnetism because Weyl was working on a classical and incomplete theory of electromagnetism that needed quantization. However, applying conformal transformation to Quantum Electrodyanamics would also become problematic as there is still no accepted geometric theory of Quantum Mechanics. Furthermore, the metric tensor $g_{\mu\nu}$ does not have a fundamental role in the mathematical formalism of Quantum Mechanics. Another possible reason why Weyl transformation failed in Quantum Mechanics is perhaps Weyl\rq{}s original conformal transformation needs to be modified in order to be integrated in the mathematical formalism of Quantum Mechanics which involves the use of complex numbers. Historically, the success of Weyl transformation in modifying General Relativity, motivated London, Fock and Weyl to use a similar transformation in Quantum Mechanics\cite{Weylhis}. They turned Weyl\rq{}s conformal (or \lq{}\lq{}gauge") transformation into a local phase transformation by changing the conformal (or scale) factor from a real function into a complex function $e^{i\alpha}$ and applying it to the wave function $\psi$ instead of the metric tensor,i.e.,
%\mathcenter
\begin{equation}
\psi\rightarrow \psi\rq{}= e^{i\alpha}\psi
\end{equation}
\noindent for some scalar function $\alpha$. For our theory, we also change the conformal factor into a complex function, but we retain the metric tensor, i.e.,
\begin{equation}\label{wgt}
g_{\mu\nu}\rightarrow \tilde{g}_{\mu\nu}=\varphi^{2}g_{\mu\nu}
\end{equation}
\noindent for some complex function $\varphi=e^{i2\pi\chi}$ written in terms of a scalar function $\chi$. In applying Weyl transformation at Planck scale, it is important first to have the necessary Physics at Planck scale. Particularly, we wanted to know the Physics of spacetime at the Planck scale. At present, it is safe to say that no one knows the Physics in such region. Though, in recent years, there are many theories that had been put forward that attempt to describe what happens at Planck scale, most of them are phenomenological that only enumerates what are the possible scenario at Planck scale and what data can be derived from an experiment at low-energy approximation. Furthermore, a complete description of the Physics at the Planck scale still remains elusive. Thus, in this paper, we will consider all the assumptions about the Planck scale that had been proposed in the past. See for example \cite{heus,ame1,Fred1,kowg,judevis,Elz09,JaRe94,Cam03,Gola08} and references therein.\\
\indent  In Postulate 2, the significant role of the second-order Ricci Flow would be in the formulation of a new mathematical formalism of Quantum Mechanics. Our approach is basically to use a normalized and conformally invariant second--order Ricci Flow to derive a new mathematical formalism of Quantum Mechanics (QM). By postulating a complex form of Weyl transformation as a fundamental symmetry in nature, it introduces the use of a conformal function $\varphi$ to modify the metric tensor $g_{\mu\nu}$. In contrast, the conventional quantum formalism (CQF) postulates the use of a wave function or a probability amplitude $\psi$ alone as a fundamental function, but no where it uses the metric tensor in a fundamental way. The main reason why the metric tensor is usually excluded in CQF, is because the role of gravity was always considered negligible at the microscopic world. Here, we suggest that the metric tensor must also have a fundamental role for a complete description of a quantum system as any quantum system will always have an inherent energy fluctuations that will lead for spacetime to fluctuate. This fluctuation of spacetime must be connected to the fluctuation of $\psi$, expressed by the famous Schr$\ddot{o}$dinger Equation or Dirac Equation. Here, in addition to the said (matter) wave fluctuation of QM, we also postulates the variation of the metric tensor in time to represent the spacetime fluctuation at the Planck scale.  In our formulation, time will remain a background parameter even at the Planck scale.
% % % % % % % % % % %     
\section*{On Postulate 1: Emergent Quantum Kinematics}\label{sec:2} 
\indent In this section, we wanted to evaluate the conformal function $\varphi$ in Eq.(\ref{wgt}) and show how it can be linked to Quantum Mechanics based on our earlier work on Deformed Special Relativity\cite{Aba2}. To do this, it is first necessary to establish the connection of Quantum Mechanics with what we called as the \lq{}\lq{}Physics at the Planck Scale". By this we mean, we will be using some of the well-known assumptions about the Planck Scale that are currently being considered by so many authors \cite{Pope00,WaBiMe06,BoWaMeBi09,For05,BeKli07,Cam02,ame,pos1,pos2,maz} such as: the existence of minimum energy and length scale, variation of fundamental constants, spacetime fluctuation and modification of Relativity and Quantum Mechanics. \\
\indent Applying now the Weyl transformation given by Eq.(\ref{wgt}) on energy dispersion relation, it gives us
\begin{align}
g_{\mu\nu}p^{\mu}p^{\nu}\rightarrow\varphi^{2}g_{\mu\nu}p^{\mu}p^{\nu}=g_{00}\tilde{E}^{2}-g_{ij}\tilde{p}^{i}\tilde{p}^{j}
\end{align}
\noindent where $\tilde{E}=\varphi E$, $\tilde{p}^{i}=\varphi p^{i}$ and $\tilde{p}^{j}=\varphi p^{j}$ are complex energy and momenta. Evaluating the complex energy and momentum, we use $i2\pi\varphi=\frac{\partial \varphi}{\partial \chi}$, $E=-\frac{\partial S}{\partial t}$, $p^{1}=\frac{\partial S}{\partial x}=p_{x}$ and inserting the imaginary number $i=\sqrt{-1}$, we have
\begin{eqnarray}\label{optder2}
\varphi E&&=-\frac{i}{2\pi}(i2\pi\varphi) E =-\frac{i}{2\pi}\left(\frac{\partial\nonumber \varphi}{\partial \chi}\right)\left(-\frac{\partial S}{\partial t}\right)\\
&&=\frac{i}{2\pi}\left(\frac{\partial \varphi}{\partial t}\right)\left(\frac{1}{f}\frac{\partial S}{\partial t}\right)\\
\label{optder3}\varphi p_{x}&&=-\frac{i}{2\pi}(i2\pi\varphi)p_{x} =-\frac{i}{2\pi}\nonumber \left(\frac{\partial \varphi}{\partial \chi}\right)\left(\frac{\partial S}{\partial x}\right)\\
&&=-\frac{i}{2\pi}\left(\frac{\partial \varphi}{\partial x}\right)\left(\lambda\frac{\partial S}{\partial x}\right)
\end{eqnarray}
\noindent where $S$ is a functional that could be similar to the classical action and the following variables were defined:
\begin{align}\label{frelen}
\frac{1}{f}=\frac{\partial t}{\partial \chi}\;\;\;\;\;\text{and} \;\;\;\;\lambda=\frac{\partial x}{\partial \chi}
\end{align}
\noindent Simplifying further, we define the variable,
\begin{align}\label{mpc}
\tilde{h}= \frac{1}{f}\frac{\partial S}{\partial t}=\lambda\frac{\partial S}{\partial x}=\frac{\partial S}{\partial \chi}
\end{align}
\noindent such that from (\ref{optder2}) and (\ref{optder3}), we yield the following equations;
\begin{align}
\label{op3}\varphi E \;\;=i\tilde{\hbar}\partial_{t} \varphi\;\;\;\;\;\;\;\;\;\varphi{p}_{x}=-i\tilde{\hbar}\partial_{x}\varphi
\end{align} 
\noindent where $\partial_{x}$ and $\partial_{t}$ are partial derivatives in space and time, respectively, and $\tilde{\hbar}=\frac{\tilde{h}}{2\pi}$. Since $\varphi$ commutes with $E$ and $p_{x}$, we have the following eigenvalue equations,
\begin{align}\label{op4}
E\varphi \;\;=\hat{E} \varphi\;\;\;\;\;\;\;\;\;\; { p_{x}}\varphi= \hat{p}_{x}\varphi              
\end{align} 
\noindent which gives us an operator correspondence, 
\begin{align}
E\equiv i\tilde{\hbar}\partial_{t}=\hat{E}\;\;\;\;\;\;\;\;p_{x}\equiv -i\tilde{\hbar}\partial_{x}=\hat{p}_{x}
\end{align}   
\noindent similar to Quantum Mechanics. Lastly, we use Eq. (\ref{mpc}) and set $S=S(\chi)$. Then integrating and setting the variable $\tilde{h}$ equal to a constant $\tilde{h}_{c}$, it will yield us
\begin{align}\label{chi1}
\chi=S/\tilde{h}_{c}+\textsf{k}
\end{align}
\noindent for some  integration constant $\textsf{k}$. This will transform $\varphi$ as follows:
\begin{align}
\varphi\rightarrow\varphi_{c}= Ae^{iS/\tilde{\hbar}}
\end{align}
\noindent which is similar to quantum probability amplitude where $\tilde{\hbar}=\tilde{h}_{c}/2\pi$ is a variable and $A$ is a constant.  All of these results are equivalent to Quantum Mechanics if and only if the energy-dependent variable $\tilde{h}$ becomes constant and equal to the Planck constant. Note that the idea of a varying Planck\rq{}s \lq{}\lq{}constant" is not something new. A varying and energy-dependent Planck \lq{}\lq{}constant" was also derived by Smolin and Magueijo\cite{MaSmlet02} in their work on Deformed Special Relativity at Planck Scale. Others \cite{mang} consider it to be a time-dependent variable at Planck scale. If it is true that the Planck constant is varying at Planck Scale, it would therefore modify all known quantum-mechanical laws. Such modification may also allow for Relativity to be applicable at Planck Scale at a certain extent although it must be different compare with the usual Relativity at macroscopic scale. It is suggested here that the simple modification of quantum-mechanical equations above applies at Planck scale. Consequently, the variables $f$ and $\lambda$ will be interpreted here as quantities that describe the fundamental fluctuation of space and time at Planck scale in terms of the new variable $\chi$. We may also call $f$ and $\lambda$ as the \lq{}\lq{}frequency" and \lq{}\lq{}wavelength", respectively, of a wave that describes the spacetime fluctuation at Planck Scale. On the other hand, the variable $\tilde{h}$, represents the fundamental energy fluctuation at Planck scale in terms of $\chi$. We can relate it to $f$ and $\lambda$ by using Eq.(\ref{mpc}) to derive a modified de Broglie-Planck equations which can be expressed as follows:
\begin{equation}\label{Pladeb}
E=\tilde{h}f\;\;\;\;\;\;\;\;\;\;\;p=\frac{\tilde{h}}{\lambda}
\end{equation}
\noindent Take note that the energy $E$ is the total energy of the system. What we wanted also is to define a minimum energy scale at Planck scale. For a constant total energy $E$ in $S=\int Edt$, Eq. (\ref{chi1}) becomes,
\begin{align}
\chi=S/\tilde{h}_{c}+\textsf{k}=\frac{Et}{\tilde{h}_{c}}=\frac{E}{\tilde{h}_{c}f_{m}}=\frac{E}{E_{p}}
\end{align}  
\noindent where we set the integration contants to cancel each other, $E_{p}=\tilde{h}_cf_{m}$ and  $f_{m}=1/t$. If we define $E_{p}$ as the minimum energy, we can have a total energy $E$ in terms of $E_{p}$,
\begin{equation}
E=\chi E_{p}
\end{equation}    
\noindent Now if at Planck scale, the region in space (in one-dimension) that can be occupied by a quantum particle is in N units of minimum length, $L_{p}$, i.e.,
\begin{equation}\label{dist}
x=NL_{p}
\end{equation}  
\noindent then by Eq.(\ref{frelen}), we have,
\begin{equation}
\lambda= \frac{\partial N}{\partial\chi}L_{p}+N\frac{\partial L_{p}}{\partial\chi }\;\;\;\;\;\;\;\;\;\frac{1}{f}=\frac{\partial N}{\partial\chi}\frac{L_{p}}{v}+\frac{\partial L_{p}}{\partial \chi}\frac{N}{v}
\end{equation}
\noindent 
\noindent where $v=\frac{\partial x}{\partial t}$. If $L_{p}$ is fundamentally invariant at Planck scale and $N$ is changing, we have,  
\begin{equation}
\lambda= \frac{\partial N}{\partial\chi}L_{p}\;\;\;\;\;\;\;\;\;\;\;\;\frac{1}{f}=\frac{\partial N}{\partial\chi}\frac{L_{p}}{v}
\end{equation}
\noindent The partial derivative in the equations is related to the energy density $\rho$,
\begin{equation}\label{N}
\rho=\frac{E}{x} =\frac{\chi E_{p}}{N L_{p}}=\frac{\chi}{N}\rho_{p}
\end{equation}
\noindent where $\rho_{p}=E_{p}/L_{p}$ is the minimum energy density. For invariant $\rho_{p}$, $\frac{\partial N}{\partial \chi}=\frac{\rho_{p}}{\rho}-\frac{N}{\rho}\frac{\partial}{\partial \chi}\left(\frac{\chi}{N}\rho_{p}\right)$, which gives us,
\begin{align}
\lambda= \left[\frac{\rho_{p}}{\rho}-\frac{N}{\rho}\frac{\partial}{\partial \chi}\left(\frac{\chi}{N}\rho_{p}\right)\right ] L_{p}\;\;\;\;\;\;\;\;\;\;
f= \left[\frac{\rho_{p}}{\rho}-\frac{N}{\rho}\frac{\partial}{\partial \chi}\left(\frac{\chi}{N}\rho_{p}\right)\right ]^{-1} \frac{v}{L_{p}}
\end{align}

\vspace{25pt}
\noindent At this point, we note of the fact that the value of the energy density ratio $\frac{\rho_{p}}{\rho}$ is inherently dependent on the measurement process as any measurement process will unavoidably add an energy into the system. At present, the way by which we do our measurement process is only at low energy resolution such that anything at Planck scale can only be observed in the order of Compton scale. If we have enough energy to increase the resolution of our measurement, such that we can send a single fundamental unit of energy $E_{p}$ to observe a single unit of fundamental length $L_{p}$, then $N=1=\chi$, $\rho=\rho_{p}$, which gives us,
\begin{equation}  
\lambda = L_{p}\;\;\;\;\;\;\;\;\;\;\;\;f=\frac{v}{L_{p}}
\end{equation} 
\noindent Combining these results with equation Eq.(\ref{Pladeb}), we have,
\begin{equation}\label{mpladeb}
E= \tilde{h}\frac{v}{L_{p}}\;\;\;\;\;\;\;\;\;\;\;p=\frac{\tilde{h}}{L_{p}}
\end{equation}    
\noindent as the Planck scale equivalent of de Broglie-Planck equations which consider an invariant minimum length and a varying energy-dependent Planck \lq{}\lq{}constant". All of these results will be used and explained later in the formulation of a new quantum interpretation.\\
\indent To end this section, we note the advantage of using Weyl transformation where the conformal factor is a complex function. It allows the possibility of linking Weyl transformation to Quantum Mechanics by showing that the complex conformal term can be written as the quantum probability amplitude. Hence, we can have a piece-wise definition of a generalized conformal transformation in terms of the nature of the conformal factor, $\Omega^{2}$, i.e.,    
\begin{equation*}
  g_{\mu\nu}\rightarrow\Omega^{2}{g}_{\mu\nu}=\begin{cases}
    \varphi^{2}g_{\mu\nu} &   \text{(Planck/Quantum Scale)}    \\
     Y^{2}g_{\mu\nu} &     \text{(Classical/Macroscopic Scale)}
      \end{cases} 
\end{equation*}
\noindent where $\varphi=\varphi(\chi)=e^{i2\pi\chi}$ is a complex function in terms of the variable $\chi$ while $Y=Y(\omega)=e^{\omega}$ is a real function in terms of scalar function $\omega$. The former is postulated to be the metric tensor of the spacetime at the Planck and quantum scale while the latter is the metric tensor at the macroscopic scale. The conformal metric tensor 
\begin{align}\label{wgcov}
\bar{g}_{\mu\nu}=\psi^{2}g_{\mu\nu}
\end{align}
\noindent is suggested here as the low-energy approximation of the metric tensor $\tilde{g}_{\mu\nu}=\varphi^{2}g_{\mu\nu}$ at Planck scale where its conformal function $\varphi$ becomes the quantum probability amplitude $\psi$ or the wave function in QM. This is in line with the work of Dzhunushaliev \cite{Dzh1} where he suggested that \emph{the metric tensor can be considered as a microscopical state in a statistical system} and by Isidro et.al.\cite{IsSanCor09a} that \emph{the State Vector or the wave function is related to the conformal term of the metric tensor}. \\
\indent In summary, the results give us one of the ways by which to describe the Physics at the Planck scale, i.e., by modifying Quantum Mechanics. However, another way to describe physics at the Planck scale is by modifying General Relativity or the dynamics of spacetime by formulating a new gravity theory to describe the spacetime fluctuations that occur at the Planck scale. This will be the topic in the next section.

 \section*{On Postulate 2: Emergent Quantum Dynamics} \label{sec:3}
\subsection*{Ricci Flow Formalism}
  In the first two decades of the 21st century, tools in Differential Topology like the  Ricci Flow began to be used in theories that attempt to describe the spacetime at the Planck Scale\cite{rf1,ric1, ric2, ric3, ric4, ric5}. One of such theories is the work of Dzhunushaliev\cite{Dzh1} where he suggested that the \emph{Ricci Flow is a statistical system that can be used to describe the topology change at the Planck Scale}. In fact, Isidro et.al. \cite{IsSanCor09a,IsSanCor09b} showed that \emph{Schr$\ddot{o}$dinger Equation can be derived from a conformally flat metric under Ricci Flow}. A similar approach will be used here but a generalized form of Dirac Equation will be derived upon consideration of the invariance of the Ricci Flow under Weyl transformation. Also, instead of the usual Ricci Flow, a second-order version of the Ricci Flow will be postulated to describe the topology change at the quantum scale. This is to account for the wave nature of a quantum particle and relate it to the wave nature of spacetime curvatures.\\
 % % % % % % % % %
 \subsection*{Normalized Ricci Flow}
 \indent Ricci Flow (R.F.) is a first-order partial differential equation formulated by Hamilton\cite{Ham95} in 1982 to be used as an analytical tool to solve topological problems. In 2003, it was used by Perelman\cite{Per09,Per59} to prove Poincare's conjecture and Thurston's Geometrization conjecture\cite{thurs}. It is usually expressed as follows,
\begin{equation}\label{riccif}
\frac{\partial g_{ij}}{\partial t}=-2R_{ij}
\end{equation}
\noindent It is a means to stretching the metric tensor $g_{ij}$ from negative curvature to positive curvature or to smooth out an arbitrary Riemannian manifold to make it look more symmetric. This is from the point of view of mathematicians. For it to have a physical meaning, we relate it here to General Relativity which is possible since the Einstein Field Equation in vacuum case is derivable from a normalized Ricci Flow. For our purpose, the partial derivative in Eq.(\ref{riccif}) is considered here to be in terms of the real physical time $t$ and $R_{ij}$ is the Ricci Tensor which measures the spacetime curvature.\\
\indent Consider the following evolution equations for the Ricci tensor($R_{ij}$) and Weyl tensor($W_{ijkl}$) under R.F.\cite{cati},
%\begin{widetext}
\begin{eqnarray}\label{evoR}
\frac{\partial R_{ij}}{\partial t}&&=\nabla^{2}R_{ij}+2g^{pr}g^{qs}R_{piqj}R_{rs}-2g^{pq}R_{pi}R_{qj}\\ \nonumber
\frac{\partial W_{ijkl} }{\partial t}&&=\nabla^{2}W_{ijkl}+2(D_{ijkl}-D_{ijlk}+D_{ikjl}-D_{iljk})\\
&&-g^{pq}(R_{ip}W_{qjkl}+R_{jp}W_{iqkl}+R_{kp}W_{ijql}+R_{lp}W_{ijkq})\\\nonumber &&+\frac{1}{2}g^{pq}(R_{ip}R_{qk}g_{jl}-R_{ip}R_{ql}g_{jk}+R_{jp}R_{ql}g_{ik}-R_{jp}R_{qk}g_{il})\\\nonumber
&&-\frac{R}{2}(R_{ik}g_{jl}-R_{il}g_{jk}+R_{jl}g_{ik}-R_{jk}g_{il})+R_{ik}R_{jl}\\\nonumber
  &&-R_{jk}R_{il}+\frac{R^{2}-|Ric|^{2}}{2}(g_{ik}g_{jl}-g_{il}g_{jk})\nonumber
\end{eqnarray}
%\end{widetext}
\noindent where $D_{ijkl}=g^{pq}g^{rs}W_{pijr}W_{slkq}$\;  and 
\begin{equation}
\nabla^{2}=g^{ij}\nabla_{i}\nabla_{j}=g^{ij}\left(\frac{\partial^{2}}{\partial x^{i}\partial x^{j}}-\Gamma^{k}_{ij}\frac{\partial}{\partial x^{k}} \right)
\end{equation}
\noindent is the Laplacian with respect to the evolving metric with the ordinary affine connection $\Gamma$ in Reimannian geometry. These equations describe the evolution of the curvatures. However, what is also important for our purpose is the preservation of the volume structure and not just the evolution of spacetime curvatures. The evolution of the volume element $dv=\sqrt{(\det{g_{ij}})}dx$ of the spacetime $M$ is given by \cite{Ham95},
\begin{equation}\label{vol}
\frac{\partial}{\partial t}\log\sqrt{(\det{g_{ij}})}dx=\frac{1}{2}g^{ij}\frac{\partial}{\partial t}g_{ij}=r-R
\end{equation}
\noindent where $r=\frac{\int_{M}Rdv}{\int_{M}dv}$ is the average scalar curvature and serves to normalize the R.F., so that the volume is constant. If the volume is not constant and fluctuating in time, to prevent the solution of Eq. (\ref{vol}) from shrinking to a point or expanding to $\infty$, we must consider the Normalized Ricci Flow (NRF)\cite{Ham95}:
\begin{equation}\label{nrf}
\frac{\partial g_{ij}}{\partial t}=\frac{2}{n}rg_{ij}-2R_{ij} 
\end{equation}    
\noindent such that,
\begin{equation}
\frac{\partial}{\partial t}\log\sqrt{(\det{g_{ij}})}=0 
\end{equation}    
\noindent where $n$ is the number of dimension. However, a general case would be to consider the non-linear NRF by getting the second derivative of Eq.(\ref{nrf}) and using Eq.(\ref{evoR}),
\begin{equation}\label{nrf2}
\frac{\partial^{2} g_{ij}}{\partial t^{2}}=\frac{2}{n}r\frac{\partial g_{ij}}{\partial t}+2\nabla^{2}_{g}R_{ij}+4g^{pr}g^{qs}R_{piqj}R_{rs}-4g^{pq}R_{pi}R_{qj}\nonumber
\end{equation} 
\noindent for constant $r$. Defining a damping term $F=F\left(\frac{\partial g_{ij}}{\partial t}, g_{ij}\right)$ which is a function of $g_{ij}$ and its first order derivative, we have the following wave equation,
 \begin{equation}\label{wave10}
 \frac{\partial^{2} g_{ij}}{\partial t^{2}}-k\nabla^{2}_{g}g_{ij}+F=0
 \end{equation}
\noindent However, a more natural starting point  would be to consider a non-linear form of Ricci Flow expressed by a wave equation of the metric tensor, 
\begin{equation}
\frac{\partial^{2} g_{ij}}{\partial t^{2}}=-2R_{ij}
\end{equation}
\noindent known as the Hyperbolic Geometric Flow (HGF). It is an equation first introduced by De-Xing Kong and Kefeng Lui in 2006. We postulate that it is an equation that describe the inherent fundamental spacetime fluctuation at the Planck Scale (using Postulate 2) while a quantum field can also be represented by such spacetime fluctuation but is held together by the gravitational attraction of its own field energy in a confined region with volume that is invariant. This idea is similar to John Wheeler\rq{}s classical notion of geons\cite{geon} but we will show later how it can be integrated to Quantum Mechanics by conformal symmetry (using Postulate 1). For an invariant volume that is equal to a unit volume, Kong and Lui in \cite{DeKe} derived a normalized form of HGF (nHGF), 
\begin{align}\label{nhgf1}
a_{ij}\partial^{2}_{t}g_{ij}+b_{ij}\partial_{t}g_{ij}+c_{ij} g_{ij}=-2R_{ij}
\end{align}
\noindent where $\partial_{t}$ is the partial derative in time, while $a_{ij}, b_{ij}$ and $c_{ij}$ are certain smooth functions in spacetime $M$. Similar to Eq.(\ref{wave10}), nHGF can also be written into a wave equation but with a damping term. To show this explicitly, an approximation for the Ricci Tensor $R_{ij}$ in terms of the metric tensor \cite{ChoKno} can also be used, i.e.,
\begin{align}\label{chowk}
R_{ij}\approx -\frac{1}{2}\nabla^{2}g_{ij}
\end{align}
\noindent which will then transform the normalized Hyperbolic Geometric Flow (nHGF) as follows:
 \begin{align}\label{nhgf}
\tilde{\square}g_{\mu\nu}=0
\end{align}
\noindent where $\tilde{\square}=a_{\mu\nu}\partial_{t}^{2}-\nabla^{2}+b_{\mu\nu}\partial_{t}+c_{\mu\nu}$ is a modified d' Alembert operator and all Latin indices were replaced to Greek indices for 4-dimensional consideration. 
\\
\subsection*{nHGF as Modified General Relativity}\label{sec:mgr}

\indent The connection of nHGF to General Relativity can be set up when $a_{\mu\nu}=b_{\mu\nu}=0$ and $c_{\mu\nu}=c$, where $c$ is equal to the Ricci Scalar which gives us the vacuum case for Einstein Field Equation (EFE),
\begin{align}
R_{\mu\nu}= \frac{1}{2}cg_{\mu\nu}
\end{align}
\noindent To derive nHRF, we use the approach of Dai et.al.\cite{dai} by first defining a metric with an orthogonal time-axis,
\begin{equation}
ds^{2}=-dt^{2}+g_{ij}dx^{i}dx^{j}
\end{equation}
\noindent where time here is an independent background parameter. By substituting this to the vacuum case,
\begin{equation}
G_{\mu\nu}=0
\end{equation}
\noindent it will yield us the so-called Einstein's Hyperbolic Geometric Flow (EHGF),
 \begin{equation}\label{1}
\frac{1}{2}g^{ij}\frac{\partial^{2} g_{ij}}{\partial t^{2}}+\frac{1}{4}g^{ij}g^{kl} \frac{\partial g_{ij}}{\partial t}\frac{\partial g_{kl} }{\partial t } -\frac{1}{2}g^{ij}g^{kl}\frac{\partial g_{ik}}{\partial t}\frac{\partial g_{jl} }{\partial t}+R=0
\end{equation}
\noindent which is similar to Eqn.(\ref{nhgf1}) but with two first-order time derivatives of the metric tensor as extra terms. Now, using the ADM formalism of General Relativity wherein it defines a Hamiltonian constraint given by,
\begin{equation}\label{0}
g^{-1}\left(\frac{1}{2}g_{ij}g_{kl}-g_{ik}g_{jl}\right )\pi^{ij}\pi^{kl}+R=0   
\end{equation} 
\noindent which is known to be fully equivalent to all ten components of the vacuum EFE, we can combine, Eqn.(\ref{0}) with Eqn.(\ref{1}), to give us,
\begin{equation}
g^{-1}\left(\frac{1}{2}g_{ij}g_{kl}-g_{ik}g_{jl}\right )\pi^{ij}\pi^{kl}-\frac{1}{2}g^{ij}\frac{\partial^{2} g_{ij}}{\partial t^{2}}-\frac{1}{4}g^{ij}g^{kl} \frac{\partial g_{ij}}{\partial t}\frac{\partial g_{kl} }{\partial t } +\frac{1}{2}g^{ij}g^{kl}\frac{\partial g_{ik}}{\partial t}\frac{\partial g_{jl} }{\partial t}=0
\end{equation}
\noindent This a form of EFE in vacuum case but in terms of first and second order derivatives of the metric tensor. In essence, the equation above is a wave equation of the metric tensor that unifies Ricci Flow and EFE.

%%%%%%%%%%%%%%%%%
\subsection*{Deriving a Unified Quantum Wave Equations}\label{gde}
To show that out of the spacetime fluctuation at the Planck scale, Quantum Mechanics emerges, we apply now our modified Weyl transformation on metric tensor on the simplest form of nHGF given by Eqn.(\ref{nhgf}). It transforms nHGF as follows,
\begin{align}\label{qhgf}
\tilde{\square}g_{\mu\nu}=0\rightarrow\tilde{\square}\bar{g}_{\mu\nu}=2g_{\mu\nu}\psi\tilde{\square}\psi+ \psi^{2}\tilde{\square}g_{\mu\nu}=0
\end{align}
\noindent where the modified (complex) Weyl transformation that we used is given by Eq.(\ref{wgcov}) as a low-energy approximation of Eq.(\ref{wgt}) that we considered to apply at the Planck scale. In order for the equation above to be invariant under Weyl transformation, the first term must vanish which gives us the following equation,  
\begin{align}\label{hgfde}
\tilde{\square}\psi=a_{\mu\nu}\partial_{t}^{2}\psi-\nabla^{2}\psi+ b_{\mu\nu}\partial_{t}\psi-c_{\mu\nu}\psi=0
\end{align} 
\noindent Notice that by setting the values of $a_{\mu\nu}$, $b_{\mu\nu}$ and $c_{\mu\nu}$ as follows:
\begin{align}
a_{\mu\nu}=\textbf{1},\;\; b_{\mu\nu}=2\left(\frac{im_{0}c^{2}\boldsymbol{\beta}}{\hbar}\right),\; c_{\mu\nu}=\left(i\frac{m_{0}c^{2}\boldsymbol{\alpha}}{\hbar}\right)^{2}\nonumber
\end{align}
\noindent where $\textbf{1}$ is the identity matrix, $\boldsymbol{\alpha} = \begin{pmatrix}0\;\;\;\sigma \\ \sigma \;\;\; 0\end{pmatrix}$  is written in terms of Pauli matrices $\sigma$ and  $\boldsymbol{\beta} = \begin{pmatrix}\textbf{1}\;\;\;\;\; 0 \\ 0 \; -\textbf{1}\end{pmatrix}$, Eq.(\ref{hgfde}) will yield us the following equation:
 \begin{align}\label{sde}
\textbf{1}\partial_{t}^{2}\psi-\nabla^{2}\psi+ 2\left(\frac{im_{0}c^{2}\boldsymbol{\beta}}{\hbar}\right)\partial_{t}\psi-\left(\frac{m_{0}c^{2}\boldsymbol{\alpha}}{\hbar}\right)^{2}\psi=0
\end{align} 
%\mathcenter
\noindent which is an equation that was first derived by Arbab\cite{Arb2011feb,ArbYass2012} using a quaterion formalism of Quantum Mechanics. He called it as the \lq{}\lq{}Unified Quantum Wave Equation\rq{}\rq{} (UQWE) as he was able to show that Dirac Equation, Klein-Gordon Equation, and Schr$\ddot{\mathsf{o}}$dinger Equation can all be derived from such single equation\cite{Arb3a}. In fact, it is just a second order version of Dirac equation. From Eq.(\ref{sde}), it yields us a second-order operator 
\begin{align}
(\boldsymbol{\alpha}\cdot\nabla)^{2}= \textbf{1}\partial_{t}^{2}+ 2\left(\frac{im_{0}c^{2}\boldsymbol{\beta}}{\hbar}\right)\partial_{t}-\left(\frac{m_{0}c^{2}}{\hbar}\right)^{2}
\end{align}
\noindent where $\boldsymbol{\alpha}^{2}=\boldsymbol{\beta}^{2}=\textbf{1}^{2}=\textbf{1}$. Factoring and getting the square root will give us the following linear operator;   
\begin{align}
\boldsymbol{\alpha}\cdot\nabla=\textbf{1}\partial_{t}+\frac{im_{0}c^{2}\boldsymbol{\beta}}{\hbar}
\end{align} 
\noindent Arranging and putting back the function $\psi$, it will give us 
 \begin{align}
\textbf{1}\partial_{t}\psi-\boldsymbol{\alpha}\cdot\nabla\psi+\frac{im_{0}c^{2}\boldsymbol{\beta}}{\hbar}\psi=0
\end{align} 
\noindent which is the Dirac Equation \cite{de}. Though this is definitely not an exact derivation of Dirac Equation as values of $a_{\mu\nu}$, $b_{\mu\nu}$ and $c_{\mu\nu}$ were arbitrarily given, the derivation of a generalized UQWE is enough for our goal in this paper. More importantly, a unifying link between UQWE (a generalized and modified Dirac Equation) and nHGF (a generalized and modified Einstein Field Equation) was established as expressed by Eq.(\ref{qhgf}) to describe the spacetime fluctuation at Planck scale from which equations of Quantum Mechanics can all be derive. In fact, from UQWE other quantum equations can also be derived. These are the Proca Equation and the Quantum Hyperbolic Heat Transport Equation. Consider a charge in a quantum system which modifies the space in the vicinity as it gives additional energy into the system. In the presence of an electric charge $q$, it yields us a phase transformation in $\psi$ since the action $S$ will transform as follows:
\begin{align}\label{act}
S=\int\{T-qA^{\mu}\}dt=\int Ldt
\end{align}
\noindent where $L$ is the Lagrangian, $A^{\mu}=(A,\phi)$ is the electromagnetic 4-potential, $T$ is the kinetic energy and we use the convention $\hbar=c=1$. If we consider the average kinetic energy of the quantum system to be related to the temperature $T_{H}$ of the system via the equipartition rule: $T=\frac{1}{2}k_{b}T_{H}$, it gives us the following equation:
\begin{equation}\label{qhte}
\tilde{\square}T_{H}=a_{\mu\nu}\partial_{t}^{2}T_{H}-\nabla^{2}T_{H}+b_{\mu\nu}\partial_{t}T_{H}+c_{\mu\nu}T_{H}=0
\end{equation}
\noindent The equation above is known as the Quantum Hyperbolic Heat Transport equation \cite{lowski} in its most generalized form. This implies that a quantum particle/field also serves as a carrier of heat and by itself can be considered as a thermodynamic system. Related to this, as explicitly shown in the Lagrangian, we can derive another wave equation for the electromagnetic four-potential $A^{\mu}$, 
\begin{align}
\tilde{\square}A^{\mu}=0
\end{align} 
 \noindent This is for the general case since by expanding the equation, we have
\begin{align}
\tilde{\square}A^{\mu}=\Box A^{\mu}+ b_{\mu\nu}\partial_{0}A^{\mu}+m_{0}^{2}A^{\mu}=0
\end{align} 
\noindent where we can set $b_{\mu\nu}=2im_{0}\boldsymbol{\beta}\ne 0$ and
\begin{align}\label{lgc}
b_{\mu\nu}\partial_{0}A^{\mu}= \partial^{\mu}(\partial_{\nu}A^{\nu})
\end{align}
\noindent such that $\partial_{\nu}A^{\nu}\ne 0$ and we get
\begin{align}\label{proca}
\Box A^{\mu}+ \partial^{\mu}(\partial_{\nu}A^{\nu})+ m_{0}^{2}A^{\mu}=0
\end{align}  
\noindent which is the Proca equation for spin-1 particle with mass\cite{poe}. Similar to GR where an energy stress tensor $T_{\mu\nu}$ was added, we can add an external source-charge by adding the 4-current density $J^{\mu}=(J, \rho)$,
\begin{align}\label{kgder}
\Box A^{\mu}+ \partial^{\mu}(\partial_{\nu}A^{\nu})+m_{0}^{2}A^{\mu}=J^{\mu}
\end{align}
\noindent which becomes the inhomogenous Maxwell equation if and only $m_{0}=0$ and the Lorenz gauge condition is preserved. Take note of the fact that the generalization of the Lorenz gauge is naturally expressed in Eq.(\ref{lgc}) and integrated in the theory in order to derive the Proca Equation. 

\section*{On the Consequence of Postulate 1 and 2:\\
Topo-Metrodynamic Interpretation (TMI)}\label{sec:4}
\indent In this section, a new quantum interpretation will be put forward by analyzing the metric solution of the nHGF. However, it is not intended here to come up with a fully developed quantum interpretation. The section is just meant to be a starting point for a new approach in solving the foundational problems in Quantum mechanics. A survey of other quantum interpretations and related topics will not be done except for reference to the Copenhagen Interpretation and Bohmian Mechanics for comparison. Only topics such as entanglement and double slit experiment will be discussed within the context of the proposed quantum interpretation which we formally called here as Topo-Metrodynamic Interpretation(TMI). The focus would be on the dynamics and fundamental nature of a quantum field that can lead to the possible justification of the \lq{}\lq{}ER=EPR" conjecture.\\ 
\indent In retrospect, the root cause of all the confusion surrounding the interpretational problem of Quantum Mechanics is that no one really knows what the true nature of the so-called \lq{}\lq{}quantum wave". It all started with the double-slit experiment which seems to suggest that a quantum particle is behaving like a wave, but no one is really sure if the electron \emph{is} the wave itself or the electron is just riding or moving along with a \lq{}\lq{}guiding wave". Feynman called the problem of explaining the double-slit experiment as \lq{}\lq{}the central mystery"\cite{Feyn80} in deciphering what we are actually doing when we are using Quantum Mechanics. Without the full grasp of the physical nature of the wave that is associated with a quantum particle, Bohr and others resort to suggest that there is really no physical wave that is waving or being propagated. The wave function is said to be just a mathematical tool that encapsulates all the things that can be known about a quantum system. This pragmatic and minimalist approach of the Copenhagen Interpretation (CI) has become the favored interpretation among physicists nowadays. For CI, the wave function $\psi$ and its corresponding wave equation, should not be postulated to describe a \lq{}\lq{}quantum wave" that physically exists as what de Broglie and Bohm seems to suggest \cite{Boh52, Bell2} with their own versions of quantum interpretation. The Wave Function for CI is just a mathematical tool to describe a quantum state or to predict the possible outcome of an experiment. Here, a new interpretation for the Wave Function or a quantum field will be put forward based on the new quantum formalism presented in the previous sections. \\

\subsection*{The Quantum Wave}

\indent The so-called quantum field is suggested here to be an ensemble of waves or fluctuations that physically exist in a quantum system. As discussed in the previous section, we can have at least three types of wave that can be unified under a single equation,  
\begin{align}\label{w1aves}
\left[\int(\tilde{\square}T-q\tilde{\square}A^{\mu})dt +g^{\mu\nu}\tilde{\square}g_{\mu\nu}\right]\psi^{2}=0
\end{align}
\noindent by using Eq.(\ref{qhgf}) and Eq.(\ref{act}). The first type of wave is an \lq{}\lq{}energy wave\rq{}\rq{} and described by the wave equation, 
\begin{align}
\tilde{\square}T=a_{\mu\nu}\partial_{t}^{2}T-\nabla^{2}T+b_{\mu\nu}\partial_{t}T+c_{\mu\nu}T=0
\end{align}
\noindent It is a wave equation that describe the energy fluctuation of a quantum system. If the kinetic energy $T$ is under the equipartition rule $T=\frac{1}{2}\chi k_{b}T_{H}$ with $\chi$ number of fundamental units of energy, then we yield the Quantum Hyperbolic Heat Transport Equation,  
%\mathcenter
\begin{align}
 \tilde{\square}T_{H}=a_{\mu\nu}\partial_{t}^{2}T_{H}-\nabla^{2}T_{H}+b_{\mu\nu}\partial_{t}T_{H}+c_{\mu\nu}T_{H}=0
\end{align}

\noindent which implies the thermodynamic nature of a quantum system as a carrier of heat. For the second type of wave, it involves the fluctuation of electromagnetic vector and scalar potential. It is a fluctuation that generates the electromagnetic field. The fluctuation is described by the equation below
\begin{align}
\tilde{\square}A^{\mu}=a_{\mu\nu}\partial_{t}^{2}A^{\mu}-\nabla^{2}A^{\mu}+ b_{\mu\nu}\partial_{t}A^{\mu}+c_{\mu\nu}A^{\mu}=0
\end{align} 
%\mathcenter
\noindent which is a general case of Proca Equation in the vacuum case. In the famous Aharonov-Bohm Experiment (ABE), it is very much established that electromagnetic potentials are \lq{}\lq{}physical" and not just a convenient mathematical tool for calculating force fields. Its influence in the surrounding space is that it gives additional energy to the system. Thus, the presence of a charge (not necessarily an electric charge), adds to the energy fluctuations that is associated with a quantum field. The consequence of such energy fluctuations is that the spacetime fluctuates. The corresponding spacetime fluctuation that is associated with a quantum field must be able to confine the energy within a very small volume of space.  In fact, the corresponding wave associated with it, is a soliton in nature and can be described by the third wave equation which is the nHGF:
\begin{align}
\tilde{\square}g_{\mu\nu}=a_{\mu\nu}\partial_{t}^{2}g_{\mu\nu}-\nabla^{2}g_{\mu\nu}+ b_{\mu\nu}\partial_{t}g_{\mu\nu}+c_{\mu\nu}g_{\mu\nu}=0.\nonumber
\end{align} 
\noindent By rearranging the terms in nHGF, we can express it into a wave equation with a damping term, $\Phi$,
\begin{equation}\label{term}
\tilde{\square}g_{\mu\nu}=\square g_{\mu\nu}+\Phi=0
\end{equation}
\noindent where $\square=a_{\mu\nu}\partial^{2}_{t}-\nabla^{2}$ and the damping term $\Phi$ is given by
\begin{equation}
\Phi=b_{\mu\nu}\partial_{t}g_{\mu\nu}+c_{\mu\nu}g_{\mu\nu}
\end{equation}
\noindent This soliton-like nature can account for the confinement or localization of the energy and charge, which can be interpreted to be the origin of the particle-like behavior of a quantum field. However, it is possible that this behaviour can only happen upon interaction of the quantum field with other fields or during the act of measurement or detection, as the quantum wave of the detector or the measurement apparatus interferes (either constructively or destructively, coherently or decoherently) with the quantum wave of the particle that is being observed. Hence, in TMI, there is no Measurement Problem, only the transaction of waves that are interacting, quantum-mechanically. In empty space, quantum fields are created in a very short time, and the interaction with such quantum fields will be the topic of succeeding sections.

\subsection*{Wormhole and Black Hole Metric}\label{sec:BHN}
\indent In their paper on nHGF, Shu and Shen were able to show that nHGF satisfies Birkhoff's Theorem and were able to solve an exact metric solution of it, which turns out to be a black hole solution\cite{ShuShe}. The metric was shown to be,
\begin{align} \label{shsh}
ds^{2}= u(r) \left(1-\frac{2m}{r}-\frac{\Lambda}{3}r^{2}\right)c^{2}dt^{2}-\left (1-\frac{2m}{r} -\frac{\Lambda}{3}r^{2}\right )dr^{2}+r^{2}d\Omega^{2}
\end{align}
\noindent where  $u(r)=(r-r_{a})^{\sigma_{a}}(r-r_{b})^{\sigma_{b}}(r-r_{c})^{\sigma_{c}}$, $\sigma_{i}=\frac{\Lambda r_{i}}{\kappa_{i}}$, $\kappa_{i}$ corresponds to the surface gravity of the black hole, while $r_{a}$, $r_{b}$ and $r_{c}$ are three roots of $1-\frac{2m}{r}-\frac{\Lambda}{3}r^{2}=0$ which depends on the value of decoupled constant $\Lambda$. For $\Lambda=0$, the solution gives a  Schwarzschild metric which is well-known to have a wormhole solution that was first derived by Einstein and Rosen in their 1935 paper\cite{er}, i.e.,  
\begin{align} 
ds^{2}=\frac{\epsilon^{2}}{\epsilon^{2}+2m}c^{2}dt^{2}
- 4(\epsilon^{2}+2m)du^{2}-(\epsilon^{2}+2m)^{2}d\Omega^{2}\nonumber
\end{align}
\noindent by setting $\epsilon^{2}=r-2m$ such that the metric describes two regions of space (or sheets) for $\epsilon<0$ and $\epsilon>0$ that is joined at the hyperplane $r=2m$. For $\Lambda\ne 0$, Shu and Shen derived a black hole solution that comes in two types. The first type is the case for $\frac{1}{\Lambda}\ge \frac{9}{4}r^{2}_{g}$, where $r_{g}=2m$ is the Schwarzchild radius. This type has two horizons with radii $r_{b}$ and $r_{c}$ since $r_{a}<0$ and $r_{c}>r_{b}>0 $. It was assumed that between the two horizons, with radii $r_{c}$ and $r_{b}$, the true event horizon is the outermost horizon with radius $r_{c}$. The second type is for the case $\frac{1}{\Lambda}< \frac{9}{4}r^{2}_{g}$. For this case, Shu and Shen considered only one event horizon as the solution involves three radii in which two of them are complex ($r_{b}=r_{c}^{*}$) while the other one is real ($r_{a}$). Note however that if we set $\Lambda=\frac{3}{2}q^{2}$, for an electric charge $q$, we get the following Schwarzschild static and spherically symmetric solution for the combined field that Einstein and Rosen also derived in their 1935 paper,
%\mathcenter
\begin{eqnarray} 
ds^{2}=u(r) \left(1-\frac{2m}{r}-\frac{q^{2}}{2r^{2}}\right)c^{2}dt^{2}-\left (1-\frac{2m}{r}-\frac{q^{2}}{2r^{2}}\right )dr^{2}+r^{2}d\Omega^{2}
\end{eqnarray}
\noindent which by setting $\epsilon = r^{2}-\frac{q^{2}}{2}$ to eliminate singularities and $m=0$, the metric becomes
\begin{align} 
ds^{2}=\left(\frac{2\epsilon^{2}}{2\epsilon^{2}+q^{2}}\right)dt^{2}
-du^{2}
-\left(\epsilon^{2}+\frac{q^{2}}{2}\right)d\Omega^{2}
\end{align}
%\mathcenter
\noindent which, again, exhibits a wormhole solution. However, in 1962, John Archibald Wheeler and Robert W. Fuller published a paper \cite{whefu} showing that the type of wormhole that was derived by Einstein and Rosen, is unstable. If it connects two parts of the same universe, it will pinch off too quickly for light (or any particle moving slower than light) that falls in from one exterior region to make it to the other exterior region. May that be the case, what is clear at this point is that the theory presented here suggests that there must be a possiblity of  association of sub-microscopic black hole or wormhole for every quantum field.\\
\indent The idea that quantum fields/particles are associated with black holes at the fundamental level is not something new as it was already explored by others \cite{bh1, bh2} since the pioneering work of Carter\cite{cart}. Carter considered quantum particles as black holes with naked singularity. Here, a quantum field is considered to be a combination of fluctuating field generated by a charge that is co-oscillating or "co-waving" in time with a fluctuating space from which a sub-microscopic black hole or wormhole can emerge. This will also be the key in describing the dynamics of quantum field which is famously called \lq{}\lq{}quantum jumps and tunneling" but historically not given exact details on its very mechanism. Here, we posit the possibility of "understanding Quantum Mechanics" despite the famous warning of Richard Feynman by using the concept of "quantum wormholes" which will be developed in suceeding sections.     

\subsection*{Entropic Traversable Wormholes}
\begin{figure} \centering                                                                                      \includegraphics[width=110mm,scale=2]{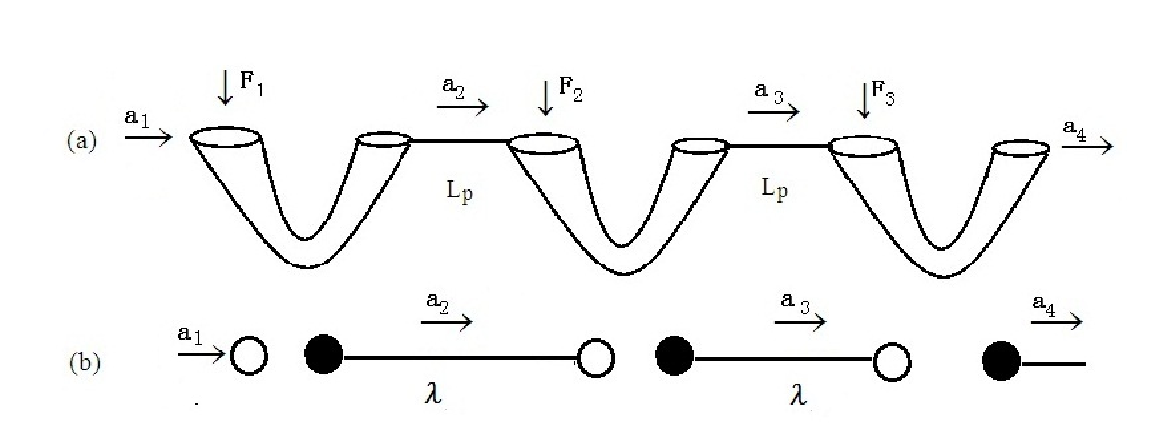}
           \caption{\textbf{Entropic Topo-Metrodynamics.} \emph{In (a), a quantum particle/field with an initial acceleration $a_{1}$ generates heat in the form of Hawking-Unruh radiation. Then the corresponding entropic force $F_{1}$ builds up a negative energy that turns the associated sub-microscopic wormhole of the quantum field to open up. The process repeats itself since the quantum field is continously interacting with its surrounding empty space such that the quantum field with its energy and charge field \lq{}\lq{}travels" by passage through a series of wormholes. As the quantum field emerges from one mouth of the wormhole to another, the minimum length it can travel is in the order of Planck length $L_{p}$. In (b), the Planck length when observed at Compton scale, is in the order of de broglie wavelength $\lambda$ as a low-energy approximation at the microscopic level. At one end or mouth of the wormhole (represented by the white circle) the quantum particle enters which to appear again at another point in space (represented by the black circle) as it emerges from the other end of the wormhole. This is the so-called \lq{}\lq{}quantum jump". The network of mouths of Planckian wormholes can be use to define the spacetime at quantum scale as an emergent spacetime.}}
           \label{fig1}
    \end{figure}

\indent In this section, we discuss another way by which a quantum field interacts with its surrounding empty space such that its associated sub-microscopic black hole turns into a traversable wormhole. Instead of using the metric solution of Shu and Shen, we can use the black hole solution of the modified EFE that we derived in Section \ref{sec:mgr}, i.e.,  
\begin{align}\label{repmefe}
G_{\mu\nu}=kT_{\mu\nu}+W
\end{align}
\noindent where the term $W=-(a_{\mu\nu}\partial_{t}^{2}g_{\mu\nu}+b_{\mu\nu}\partial_{t} g_{\mu\nu})$, is considered here to be associated with the inherent (negative) energy of empty space due to quantum fluctuations. A similar modification of EFE was also done by Novikov \cite{nov} which also involves additional terms that contain second and first order derivative of the metric tensor $g_{\mu\nu}$ with particular values for $a_{\mu\nu}$ and $b_{\mu\nu}$. He called it as \lq{}\lq{}Quantum Modification of General Relativity" as the extra terms added in EFE represent the quantum effects of the production of matter/energy by the vacuum or empty space. Another approach is to include an exotic matter content in the energy  tensor $T_{\mu\nu}$ which can be expressed as an anisotropic fluid,
\begin{equation}
T_{\mu\nu}=(\rho+p_{t})u_{\mu}u_{\nu}-p_{t}g_{\mu\nu}+(p_{r}-p_{t})x_{\mu}x_{\nu}
\end{equation} 
\noindent where $\rho$ is the energy density, $p_{r}$ and $p_{t}$ are the radial and tangential pressure, respectively, while $u_{\mu}$ and $x_{\mu}$ are the 4-vector velocities along the traverse and tangential directions which satisfy the relations $u_{\mu}u^{\mu}=-1$ and $x_{\mu}x^{\mu}=1$. These types of wormhole is what we called here as \lq{}\lq{}Planckian wormholes" where quantum consideration will allow enough time for it to be traversable upon interaction with the negative energy generated by the surrounding empty space. \\
\indent In the past, different models of traversable wormhole had been put forward that consider the role of quantum fluctuations. Most prominent of these models can be found in the work of Morris and Thorne \cite{moth} where they proposed traversable wormholes as solutions of Einstein's Field Equation that contain exotic matter  with  negative  energy  density. The stability of the wormhole depends on the exotic matter content. Classically, it is not possible to have enough negative energy sources as it violates the Average Null Energy Condition (ANEC). Consequently, there are also many attempts to have a model of traversable wormhole which not necessarily needed an exotic matter or negative energy sources like the wormhole models of Visser\cite{vis}. All of this however is at the macroscopic level. At Planck scale, one must consider quantum effect and the only known pure quantum effect to produce negative energy is the so-called Casimir effect. In Casimir effect, the Casimir force is given by\cite{MoSt97},
\begin{equation}
F_{c}=-\frac{\pi^{2}}{240d^{4}}A_{c}
\end{equation}
\noindent where $d$ is the distance of separation between the plates and $A_{c}$ is the surface area of the plates. This force exists because of the negative energy density between the plates\cite{Kr94},
\begin{equation}
\rho_{c}=-\frac{\pi^{2}}{720}\frac{1}{d^{4}}
\end{equation}
\noindent which is associated with the zero point energy of quantum electrodynamics vacuum. The pressure can be derived by renormalizing the negative energy,
\begin{equation}
p_{c}=\frac{F_{c}}{A_{c}}=-\frac{\pi^{2}}{240}\frac{1}{d^{4}}
\end{equation}
\noindent Results similar to this can be achieved if fundamentally we treat a quantum particle as an energy and charge field that is confined within a sub-microscopic black hole. We can use some of the well-known theorems in Black Hole Physics. One of which is the idea that all black holes are thermodynamic systems that obey the 1st Law of Thermodynamics,
\begin{equation}
Fdx=T_{H}dS_{E}
\end{equation}  
\noindent from which we can define an entropic force $F$ by using the Bekenstein Formula\cite{ver},
\begin{equation}
dS_{E}= k_{b}\frac{mc}{\hbar}dA
\end{equation}
\noindent  where $S_{E}$ is the entropy of the black hole, $k_{b}$ is the Boltzmann\rq{}s constant, $m$ is the mass of a test particle and $dx$ is its distance from the  black hole which has a horizon area $A$. The entropic force is therefore given by,
\begin{equation}\label{vereq}
F=\frac{dS_{E}}{dx}T_{H}=k_{b}\frac{mc}{\hbar} T_{H}\frac{dA}{dx}
\end{equation} 
\noindent Now, the heat generated is brought about by the motion of the quantum field. As the quantum field moves in empty space, it gains mass by interacting with the Higgs field and virtual particles. However, the interaction with virtual particles will also cause for the quantum field to change its direction randomly and therefore accelerates. The acceleration brought about by absorption of virtual particles with negative energies will cause for the quantum field (as a sub-microscopic black hole) to generate heat. It is a phenomenon known as the Hawking-Unruh effect where the corresponding temperature can be expressed in terms of the particle's acceleration\cite{unruh},
\begin{equation}\label{hawk}
T_{H}=\frac{\hbar}{2\pi ck_{b}}a
\end{equation}
\noindent where $\hbar$ is the reduced Planck constant, $a$ is the acceleration, and $c$ is the speed of light. The interaction of the quantum field with empty space leads to an increase on energy, entropy and the horizon area of the corresponding black hole. However, as soon as the corresponding heat is emitted in the form of Hawking-Unruh radiation, the horizon area decreases ($\frac{dA}{dx}<0$) and the black hole returns to its initial horizon area $A_{0}$. If we set $\frac{dA}{dx}=k_{0}A_{0}$, i.e., the change in the horizon area is proportional to the initial horizon area for some constant of proportionality $k_{0}$. This will yield us an expression for the entropic force given by 
\begin{equation}
F=-k_{0}k_{b}\frac{mc}{\hbar} T_{H}A_{0}=-k_{0}\frac{ma}{2\pi}A_{0}
\end{equation}
\noindent where we substitute the value of $T_{H}$ from Eq.(\ref{hawk}). Thus, the corresponding pressure is given by,
\begin{equation}
p=\frac{F}{A_{0}}=-k_{0}\frac{ma}{2\pi}
\end{equation}
\noindent We note the striking similarity in nature of the derived entropic force and the so-called Casimir force. With this result, we ought to suggest that the type of traversable wormhole that can be associated to a quantum field, would be those considered as a Casimir type of wormhole similar to those described by Garattini \cite{garat}. This is due to the unique interaction of the corresponding sub-microscopic black hole that represents a quantum field, to its surrounding empty space. We can picture the surface of the sub-microscopic black hole that we associated with a quantum field, to be a place where there is always a build up of negative energy. Such build up of negative energy triggers the opening of a traversable wormhole for a brief period of time.\\
\indent In retrospect, the corresponding field of entropic force reminds us of Bohm's idea of \lq{}\lq{}quantum potential"\cite{Boh52} or de Broglie's concept of \lq{}\lq{}pilot wave" \cite{Bell1,Bell2} as the quantum particle interacts with its own field of entropic force generated by its own heat. It is suggested here to be the root cause of the probabilistic nature of quantum particle's motion. The motion is suggested here to be similar with Brownian motion but under the influence of field of  entropic force that act like a kind of a \lq{}\lq{}sub-quantum medium" as inferred by others \cite{ViBo54, Ma09}. However, as we have shown here, the probabilistic nature of Quantum Mechanics is not just thermodynamic in nature but also \lq{}\lq{}geometrodynamic". By this we mean that the structure of spacetime or its curvature and fluctuation also affects the \lq{}\lq{}motion" of a quantum field. Following the suggestion of Wheeler, the structure of spacetime at Planck scale must be \lq{}\lq{}foam-like" in nature due to quantum fluctuation. The background spacetime of a quantum particle as observe at Planck scale is a dynamical place where there is a continuous generation of sub-microscopic wormholes\cite{Whe57}. This is very much consistent to what we have arrived here. As the sub-microscopic black hole associated with a quantum field continuously turns into a traversable wormhole, it will allow for the energy and charge field within it to travel from one point in space to another by passing through a series of sub-microscopic wormholes (Please see Figure \ref{fig1}.). The passage within each wormhole must be within a given period of time. Such period of time must be within the limit set by Heisenberg Uncertainty Principle. However, at Planck scale, as already suggested in the previous sections, one must consider the inclusion of a minimum length and energy scale that will generalized the Heisenberg Uncertainty Principle. In fact, this is currently being explored by people who are working on traversable Casimir wormhole. One example of it, is the very recent work of Jusufi et.al. \cite{jamil} by which they construct a traversable Casimir wormhole that uses a generalized uncertainty principle that is modified by the inclusion of a minimum length scale in the order of Planck length. In this paper, we will not go into details on what would be the suitable wormhole solution for our theory as it will disgress too much on the current topic of this paper. At this point, it is enough to say that a traversable Casimir wormhole that obeys a modified Heisenberg Uncertainty Principle is what we suggest to be the type of Planckian wormhole that can be associated with a quantum field. The modification of the Heisenberg Uncertainty Principle by inclusion of a minimum length and energy scale will be the topic of the next section. \\
\indent To end this section, we note that the fundamental minimum length $L_{p}$ is not only interpreted here as a fundamental unit of spacetime but also as the minimum distance between the two mouths of a Planckian wormhole. Comparing to other quantum gravity theories, it does not consider the notion of loops\cite{loop}, spins\cite{spin} and causal sets\cite{causal,Bom87} as the fundamental building blocks of a quantized spacetime. Here we have \lq{}\lq{}holes" and \lq{}\lq{}handles" which correspond to the \lq{}\lq{}mouths" and \lq{}\lq{}throats" of the Planckian wormholes. Thus, one of the main advantages of the present theory is that, in principle, one can formally describe the whole theory using Differential Topology.

% % % % % % % % % % %
\subsection*{Modified Uncertainty Principle}\label{sec:MUP}
At present, almost all quantum interpretations have always considered Heisenberg Uncertainty Principle (HUP) to be a fundamental principle. However, the understanding of the actual meaning of the principle is still very much problematic or unclear. One intepretation is that the nature of the measuring process by which an observer measure the properties of a quantum system is simply fundamentally limited. Another is that the origin of the uncertainties is not just limited to the observer and the measurement process but to the way Nature present itself to the observer. In addition to this ontological problem, the mechanism why HUP is observe in Nature was never really given, since, in Copenhagen Interpretation, it asserts that there is no such mechanism or underlying reason why Nature follows HUP. Here, however, we suggest that the origin of the uncertainties is brought about by the fundamental nature of spacetime at the Planck scale where in the presence of an energy or a field, it is no longer smooth and continuous but has \lq{}\lq{}holes". Each hole is like a tiny rip in a smooth fabric of spacetime which is essentially the mouth of a sub-microscopic wormhole. This consideration will greatly affect the very description of the quantum field. As already pointed out in the previous section, the black hole associated with a quantum field can become a wormhole as it interacts with its surrounding empty space. Such scenario will make the field to \lq{}\lq{}spread out" to all possible exit points of the wormholes, making the field to be observable at different points in space at the same time. This is possible since there can never be a scenario that a quantum field can be isolated and separated away from an empty space. It will always be in contact with empty space which can be a rich source of negative energy. Hence, there is always a high probability that the sub-microscopic black hole\footnote{"Sub-microscopic" is define here to be the scale ranging between quantum scale to Planck scale, which include just below the quantum scale where emergence of QM from the \lq{}\lq{}Physics of the Planck scale" is more achievable.} associated with a quantum field will always turn into a traversable wormhole. In such situation, it is but natural for a quantum field to move from one point in space to another through a series of wormholes. It is as if the quantum field is in a state of continuous quantum tunneling in space without any potential barrier. As it enters from one wormhole to another, the total distance travelled $x$ from one wormhole to another is given by Eq.(\ref{dist}) and has a corresponding uncertainty,
\begin{equation}\label{lim}
\delta x= N \delta L_{p}
\end{equation} 
\noindent where $\delta L_{p}$ is the uncertainty in measuring $L_{p}$. The uncertainty in measuring $L_{p}$ is brought about by the fact that the current nature of our measurement process will never have enough energy for it to have a resolution within the Planck scale. Substituting now Eq.(\ref{lim}) to  Eq.(\ref{mpladeb}), it yields us
\begin{equation}
\delta E= \tilde{h}\frac{v}{\delta L_{p}}=\tilde{h}\frac{Nv}{\delta x},\;\;\;\;\;\;\;\;\;\;\;\delta p=\frac{\tilde{h}}{\delta L_{p}}=\frac{N\tilde{h}}{\delta x}
\end{equation}    
\noindent For $N>0$, we get a modified Heisenberg Uncertainty Principle, 
\begin{equation}\label{varhup}
 \delta E\; \delta t \geq\tilde{h}\;\;\;\;\;\;\;\;\;\;\;\delta p \;\delta x \geq\tilde{h}
 \end{equation}
 \noindent where $\delta x$ and $\delta t=\delta x/v$ are the uncertainties in position and time, respectively, while $\delta E$ and $\delta p$ are the uncertainties in energy and momentum, respectively. At low-energy approximation where $\tilde{h}\rightarrow h$, it yields us the familiar Heisenberg Uncertainty Principle. In one sense, this is similar to Bohr\rq{}s Correspondence Principle. However, instead of the behavior of systems described by Quantum Mechanics reproducing Classical Physics, it is the behaviour of systems at Planck scale that is reproducing quantum-mechanical phenomena. We note, however, that the modified HUP that we derived is bordering in the region that we define here to be \lq{}\lq{}sub-microscopic", i.e., the range just below microscopic but still far from the Planck scale. With such a very large gap, at the Planck region, the derivation of a modified HUP will need more than just a time-varying energy-dependent Planck constant. In the work of Licata et.al.\cite{lic}, they consider a more generalized and detailed modification of HUP based on various quantum gravity theories. They use,
\begin{equation}    
\Delta x\Delta p\geq \hbar\left [\frac{1}{2} +\beta l_{p}^{2}\frac{\gamma^{2}M^{2}}{2\alpha\hbar^{2}}  M^{2}_{pl}c^{2}\right ]
\end{equation}
\noindent where $M_{pl}$, is the Planck mass, $l_{p}$ is the Planck length, $\alpha$ is the fine structure constant, $M$ is an adimensional (variable) mass of the virtual particles of the \lq{}\lq{}lattice" of network of wormholes, $\gamma$ corresponds to the ratio of gravitational and electromagnetic interactions and $\beta$ is a fluctuating quantity which expresses the fact that space-time fluctuations fix the minimal scale only on average, in analogy with what happens in quantum foam scenarios. If we use the definition for $\alpha=2\pi k_{c}e^{2}/\hbar c$, we yield,
\begin{equation}    
\Delta x\Delta p\geq \left [\frac{h}{4\pi} +\beta l_{p}^{2}\frac{\gamma^{2}M^{2}}{4\pi k_{c}e^{2}}  M^{2}_{pl}c^{3}\right ]=\tilde{h}
\end{equation}
\noindent where $k_{c}$ is the Coulomb constant, $e$ is the fundamental charge. The second term inside the square bracket served as the \lq{}\lq{}fluctuation term" yielding us also a varying Planck \lq{}\lq{}constant", $\tilde{h}$ similar to Eqn. (\ref{varhup}).
% % % %

% % % % % % % % % % % % % %
    
\subsection*{The \lq{}\lq{}ER=EPR" Conjecture}
Historically, quantum entanglement was realized as a fundamental quantum property upon resolution of the famous Einstein-Podolsky-Rosen(EPR) paradox by a series of experimental confirmation in the 1980s of the violation of Bell's Inequality. One of the recently proposed ideas to explain quantum entanglement is the so-called \lq{}\lq{}ER=EPR" conjencture of Susskind and Maldacena\cite{susmal}. The conjencture posit the idea that quantum particles are entangled because it is physically connected via an Einstein-Rosen (ER) bridges or wormholes. To prove it here in the context of TMI, we use Hawking's Area Theorem which states that the horizon area of a black hole never decreases. It implies an inequality between the initial and final horizon area of the black hole, say for two merging black holes with Horizon Area $A_{1}$ and $A_{2}$, we have
\begin{equation}
A_{1}+A_{2}< A_{tot}
\end{equation} 
\noindent where $A_{tot}$ is the resulting horizon area. If the Holographic Principle holds, then the number of bits (units of information) on the horizon area is given by
\begin{equation}
\chi=\frac{k_{b}c^{3}}{4G\hbar}A
\end{equation}
\noindent where each bit corresponds to a unit of energy as a measure of entropy. However, since we are looking into processes at Planck scale region, a more appropriate measure of entropy will be the entanglement entropy based on a generalized Bekenstein's Area Theorem formula in the context of CFT using the AdS/CFT correspondence:
\begin{equation}\label{ryu}
S_{A}=\frac{Area\; of \;\gamma_{A}}{4G_{N}^{d+2}}
\end{equation} 
\noindent Eqn.(\ref{ryu}) is the famous Ryu-Takayanagi formula. In their original paper\cite{ryu}, they consider a subsystem A which consists of multiple disjoint intervals as follows
\begin{equation}\label{reg}
A = \{x|x \in [r_{1}, s_{1}] \cup [r_{2}, s_{2}] \cup \ldots \cup [r_{N} , s_{N}]\}
\end{equation}
\noindent where $ 0\leq r_{1}< s_{1}< r_{2} < s_{2} <\ldots < r_{N} < s_{N} \leq L$.
In the dual $AdS_{3}$ description, the region given in (\ref{reg}) corresponds to $\Theta \in \bigcup^{N}_{i=1} \left [\frac{2\pi r_{i}}{L} , \frac{2\pi s_{i}}{L}\right ]$ at the boundary. In terms of the disjoint intervals, $A_{i}$, for each of the square bracket in the equation above, $A=\bigcup_{i}A_{i}$,  with $A_{i}\{x\in \mathbb{R}|x\in (r_{i}, s_{i} ) \}$. We can consider geodesics that connect the left endpoint of one interval, say $A_{i}$, with the right endpoint of any other $A_{j}$ (including itself). The length of such geodesics is simply proportional to $2\log \frac{|r_{i}-s_{j}|}{\epsilon}$ where $\epsilon$ is a UV regulator\cite{fuk}. Then, using again the Ryu-Takayanagi formula, the entropy  will be,
\begin{equation}
S_{A}=min\left(\frac{c}{3}\sum\limits_{(i,j)} \log\frac{|r_{i}-s_{j}|}{\epsilon} \right)
\end{equation}
\noindent where the sum runs over all possible pairs from which we choose the globally minimum result, and $c$ is the central charge. Calculating the minimal (geodesic) lines in order to know the entropy is not straightforward. For two intervals, fro example, we have\cite{fuk},
\begin{equation}
S_{A}=min\left(\frac{c}{3}\log\frac{|r_{1}-s_{1}|}{\epsilon}+\frac{c}{3}\log\frac{|r_{2}-s_{2}|}{\epsilon},\frac{c}{3}\log\frac{|r_{1}-s_{2}|}{\epsilon}+\frac{c}{3}\log\frac{|r_{2}-s_{1}|}{\epsilon} \right)\nonumber
\end{equation}
\noindent Now, there is a known way to calculate the entropy in this region by conformal mapping\cite{cal}. But in the work of Ryu and Takayanagi, from their formula, they get the entropy to be, 
\begin{equation}\label{reg1}
S_{A} =\frac{\sum_{i,j} L_{r_{j},s_{i}}-\sum_{i< j} L_{r_{j}, r_{i}}-\sum_{i< j} L_{s_{j}, s_{i}}}{4G^{(3)}_{N}}
\end{equation}
\noindent where $L_{a,b}$ is the geodesic distance between two boundary
points $a$ and $b$, and the correct definition of the minimal surface was suggested to be given by the numerator in Eq. (\ref{reg1}). In our model, the metric tensor of the geodesic is defined by the conformal factor $\varphi^{2}$. Since there is an inequality of conformal factors,
 \begin{equation}
 \varphi^{2}_{1}+\varphi^{2}_{2}< \varphi^{2}_{tot}\end{equation} 
 \noindent which came from the notion of superposition,
\begin{align}\label{entang1}
(\varphi_{tot})^{2}=(\varphi_{1}+\varphi_{2})^{2} =\varphi_{1}^{2}+\varphi_{2}^{2}+m(A_{1},A_{2})
\end{align} 
%\mathcenter
\noindent where $m(A_{1},A_{2})$ is set as the interference term between two conjoined regions $A_{1}$ and $A_{2}$ in space. Multiplying Eq.(\ref{entang1}) with $g_{\mu\nu}$, a conformal metric tensor can be defined as a linear combination of three conformal metric tensors,
 \begin{equation}
 \bar{g}^{(tot)}_{\mu\nu}=\bar{g}_{\mu\nu}^{(1)}+\bar{g}_{\mu\nu}^{(2)}+\bar{g}^{m}_{\mu\nu}
 \end{equation}         
\noindent where $\bar{g}_{\mu\nu}^{(1)}=\varphi_{1}^{2}g_{\mu\nu}$, $\bar{g}_{\mu\nu}^{(2)}=\varphi_{2}^{2}g_{\mu\nu}$, and $\bar{g}^{m}_{\mu\nu}=m(A_{1},A_{2})g_{\mu\nu}$. This gives us,
 \begin{equation}
 \tilde{\square}\bar{g}^{(tot)}_{\mu\nu}=\tilde{\square}\bar{g}_{\mu\nu}^{(1)}+\tilde{\square}\bar{g}_{\mu\nu}^{(2)}+\tilde{\square}\bar{g}^{m}_{\mu\nu}=0
 \end{equation}
\noindent In TMI, the equation above can be interpreted to represent two quantum fields in two different regions that are in superposition with each other where each of its fluctuating spacetime is represented by $\tilde{\square}\bar{g}_{\mu\nu}^{(1)}=0$, and $\tilde{\square}\bar{g}_{\mu\nu}^{(2)}=0$, respectively. The two regions can be connected via the term $\tilde{\square}\bar{g}^{m}_{\mu\nu}=0$, which we have shown to have a wormhole solution. Thus, in essence, this is the famous \lq{}\lq{}ER=EPR" conjecture, where the mechanism that makes the properties of entangled quantum fields connected (or correlated) is via the use of a wormhole or ER bridges. In the next section, we consider the case where all these ER bridges are interconnected to one another and can be used for transfering and redirecting all the quantum properties of the quantum field in a much larger region of space.
% % % % % % % % % %

   \subsection*{The Interference Pattern}
\begin{figure} \centering                                                                          
                  \includegraphics[width=100mm,scale=1.0]{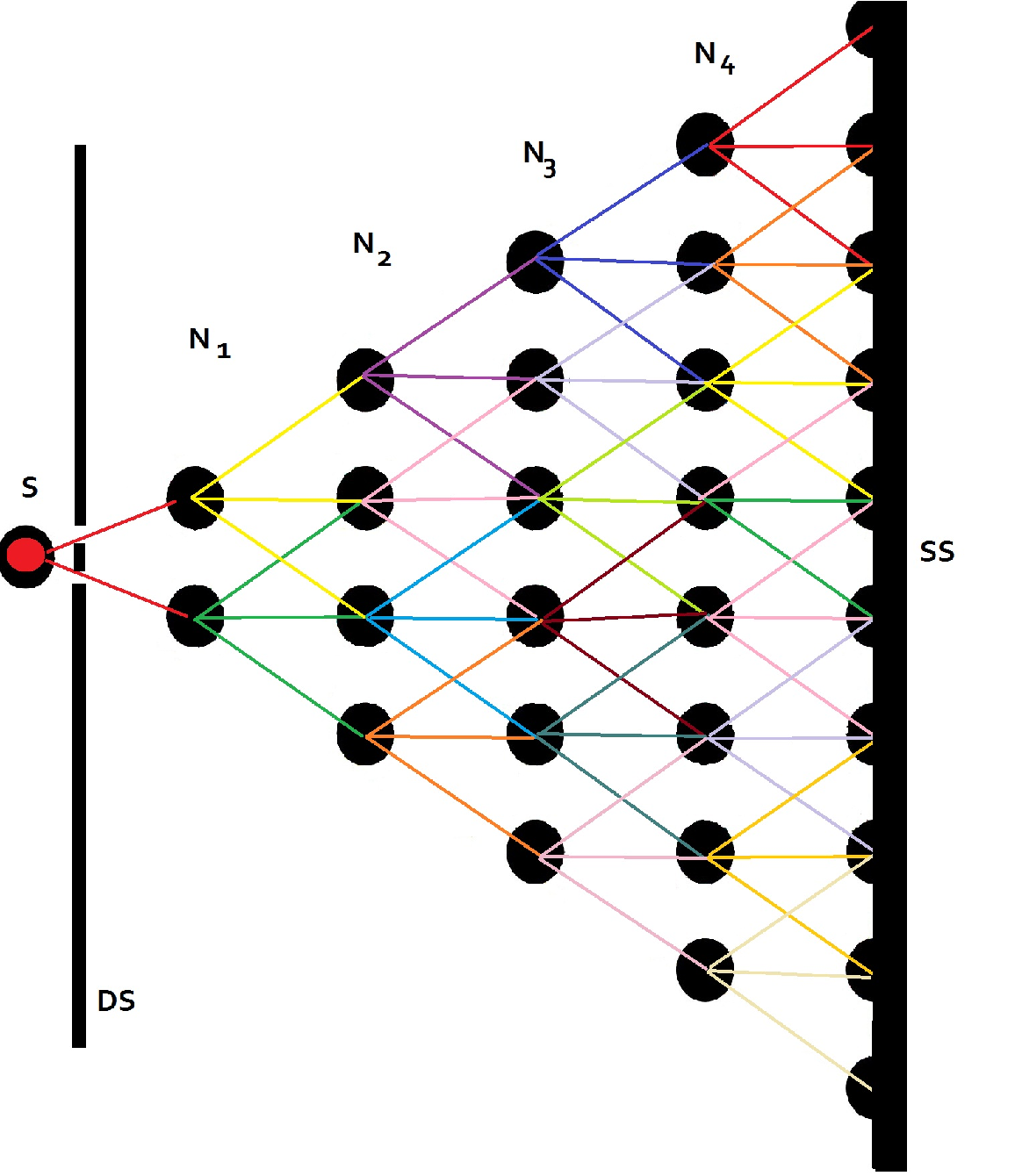}
      \caption{\textbf{A 2-dimensional representation of the double-slit experiment as an NP-hard Combinatorial Optimization Problem} \emph{From the particle source, S, a quantum particle does not necessarily pass through the double slits, DS. It can simply emerge beyond the slits through a wormhole. Then it travels from one womhole to another and emerge at every mouth (represented by the black dots, exaggerated in size) of the wormholes that it enters. The distance travelled by the quantum particle from the slits towards the scintillation screen, SS, is given by the N number of minimum length between the mouths of the wormhole. It also measures the number of possible points (or mouths of the wormholes) where the quantum particle will come out. The colored lines are all the possible path that the quantum particle can take, as N increases in time. The probability that the particle will be found to emerge at the mouth of the wormhole corresponding to $N_{i}$ units of minimum length is given by $|\psi_{i}|^{2}$ where $\psi_{i}=\psi_{i}(N_{i})$. For simplicity of the diagram, the black dots represent only the mouths where the quantum particle will emerge and the mouths were it enters are not included. Also, the possible effects on the resulting pattern made by those particles that interact with the slits are not considered. One of these effects is focusing the path of most of the particles towards the center of the scintillation screen (See Figure \ref{fig3}.)}}
      \label{fig2}
               \end{figure}

\begin{figure} \centering
                                                                         
              \includegraphics[width=100mm,scale=0.9]{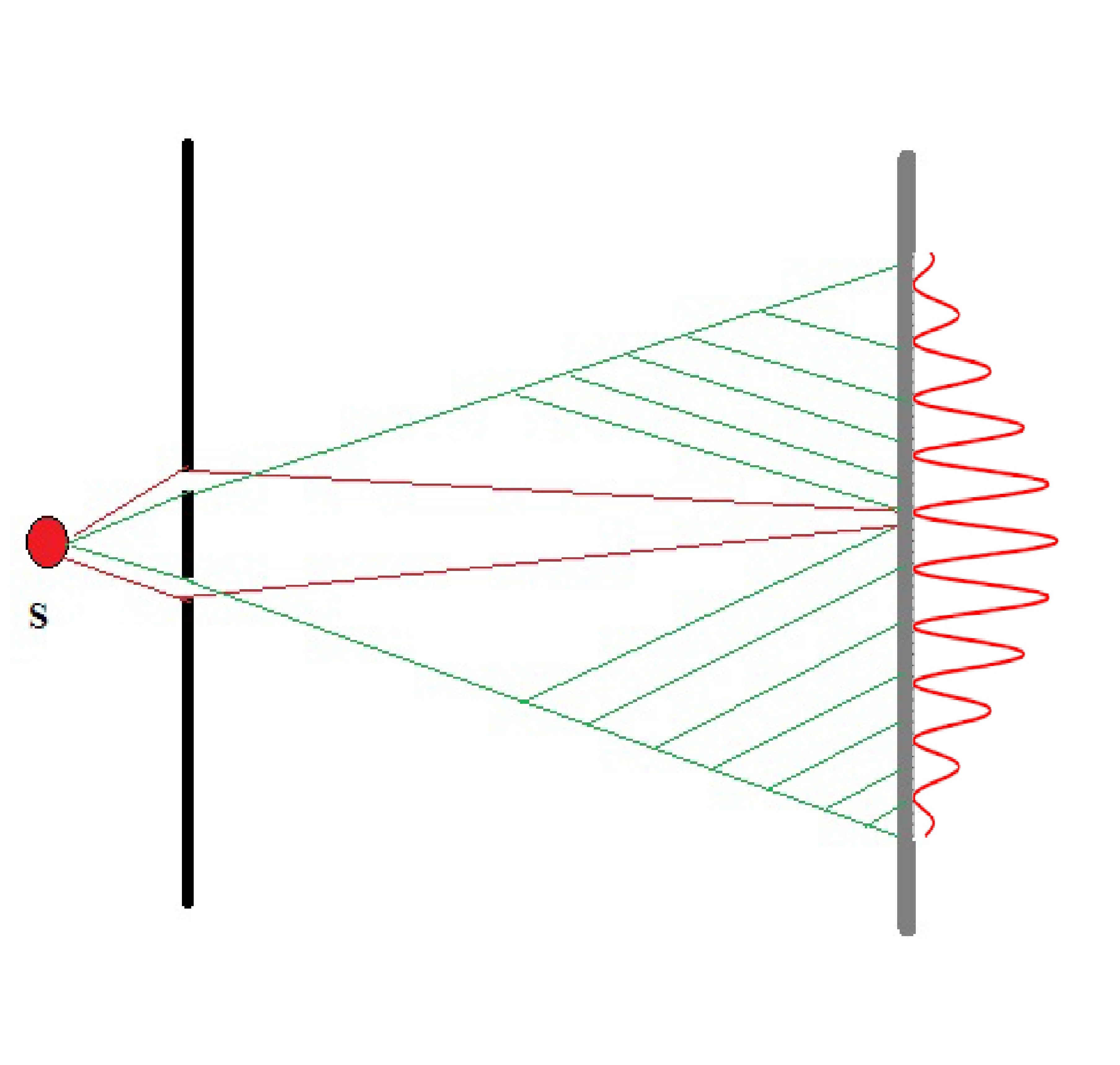}
     \caption{\textbf{Focusing Effect of the Particle's Interaction with the Slits.} \emph{When a particle interacts that constitutes either one of the two outermost edges of the slits at a certain angle, its possible path (as it traverses a series of Planckian wormholes) would bend toward the center of the detector-screen. Such paths are represented by the two violet lines at the center. On the other hand, when the particle interacts with the particle (or atom) of the innermost edges of the slits at a certain angle, the path bends toward the outermost part of the detector screen. Since it takes a longer time to reach such a part of the screen, the probability of the particle going to the center of the screen also increases. The possible paths for such a scenario are represented by a series of green lines. For simplicity, the mouths of the Planckian wormholes where the particle emerges when it exhibits the so-called \lq{}\lq{}quantum jumps" are not shown like in Fig.\ref{fig2}. Also, for simplicity of the diagram, the possible divergence of quantum jumps was not depicted (in its very small details), but only a diagonal green line (averaging) towards the center was shown. The vertical sinusoidal red line represents the graph of the frequency of quantum particles hitting a particular part of the detector screen, producing the interference pattern.} }
     \label{fig3}
            \end{figure}

  \indent As already mentioned in the previous subsections, a quantum particle/field in motion will appear to emerge from one point in space to another corresponding to every mouth of the wormhole it enters. This would easily explain the phenomenon of  \lq{}\lq{}quantum tunneling" where particle seems to emerge beyond a potential barrier. Similarly, the famous double slit experiment can also be explained where a quantum particle seems to pass thru both the slits, when in fact, it is also possible that it does not pass through at all in any of the slits. In TMI, the quantum particle simply emerges from one point in space to another since the two regions or points in space, before and beyond the slits, are connected by a wormhole. The number of points in space or mouths of the wormholes, where the quantum field enters and emerges is given by $2N$ where $N$ is the number of minimum length $L_{p}$ or the number of wormholes it enters. In Eq.(\ref{N}), the variable $N$ was shown to be proportional to $\chi$ or the number of fundamental unit of energy, i.e.,
  \begin{equation}\label{num}
               N=\left(\frac{\rho_{p}}{\rho}\right)\chi
               \end{equation}
    \noindent where, on the average, the ratio of energy densities, $\rho_{p}/\rho$ is constant if the quantum system remains isolated from any external energy source such as a measuring device, a barrier like the edge of a slit and a detector like a simple scintillation screen. For the case of a quantum particle in motion, the variable $N$ changes in time. The total distance $d$ it travelled is given by N units of $L_{p}$ while its velocity $v$ is proportional to the time derivative of $N$, 
               \begin{equation}
               v=\frac{\partial d}{\partial t}=L_{p}\frac{\partial N}{\partial t}
               \end{equation}
               \noindent This is not  the usual classical motion since the velocity is dictated by the time derivative of $N$, which can also be expressed in terms of the entropy $S_{E}$ of the system,
               \begin{equation}
               \frac{\partial N}{\partial t}=\left(\frac{\rho_{p}}{\rho}\right)\frac{\partial\chi}{\partial t}=\left(\frac{T_{H}}{Q}\frac{\rho_{p}}{\rho}\right)\frac{\partial S_{E}}{\partial t}
               \end{equation}
               
               \noindent since the number of fundamental unit of energy $\chi$ is related to the entropy $S_{E}$ of the system,i.e.,  
               \begin{equation}
               S_{E}=\frac{E_{H}}{T_{H}}=\frac{Q}{T_{H}}\chi 
               \end{equation}
               \noindent where $E_{H}$ is the total heat of the system, which is equal to $\chi$ units of the fundamental unit of heat $Q$ that is associated with each unit of fundamental energy.  Since by the Second Law of Thermodynamics, $\frac{\partial S_{E}}{\partial t}>0$, and using Eqn. (\ref{num}),
               \begin{equation}
               \frac{\partial N}{\partial t}>0
               \end{equation}    
    \noindent  Consequently, as $N$ increases in time, at each point in time, the possible points (i.e., mouth of the Casimir wormholes) in space where the quantum particle/field will emerge increase. This increase in the number of \lq{}\lq{}point of exit" from the throat, dictates the path of the particle and could form a pattern that is similar to an interference pattern if a detector screen is encountered (See Figure \ref{fig2}). Such pattern can also be seen even for a single slit, but becomes apparent in the case of double slits. There is such difference because one must also consider the case when the particle interacts with the slits. A single slit can affect the trajectory of the particle where it tends to focus the path of the particle towards the center of detecting screen. At a given angle of interaction at the edge of the slit, the particles will bound to go towards the center of the screen. Adding a second slit will add to such \lq{}\lq{}focusing effect" of the slits. The outermost edges of the slits guide the particle towards the center of the screen similar with having a single slit. On the other hand, the interaction of the particle with the innermost edges of the slits at a certain angle guides the particle toward the outer most part of the screen. Since it will take a much longer time to reach that part of the screen, the particle will interact more with its surrounding empty space and the probability of the particle going towards the center (i.e., through a series of wormholes), increases, instead of being equal to the probability of the particle going towards the outermost part of the screen (See Figure \ref{fig3}). \\
 \indent In a nutshell, if a quantum particle is a free particle without any external force or condition directing it to move in a particular direction, it will randomly go to any direction, like in a Brownian motion. In this case, the interaction with \lq{}\lq{}ordinary" virtual particles that appear in the surrounding space could be the main factor in its motion. However, under certain conditions, that the particle will also have to obey The Principle of Least Action (with all the conservation laws attached with it) and pass through a series of wormhole that open up due to "exotic" virtual particles, the motion now is similar to a Travelling Salesman. Thus in TMI, the motion of a quantum particle is not dictated by some mysterious \lq{}\lq{}guiding wave", but simply obeys the Second Law of Thermodynamics and The Principle of Least Action while it travels through a series of sub-microscopic wormholes in the background space. The problem now in describing how quantum particles go from point to another, is similar to the Travelling Salesman Problem, i.e., "Given all the possible points (or wormhole mouths) and the distances between each pair of points, what is the shortest possible route that a quantum particle can traverse to reach its destination?" \emph{It is therefore an NP-hard problem in combinatorial optimization}-a problem related to the famous "P vs NP" problem. Furthermore, to add with all the complications, there must be an inherent randomness in the generation of \lq{}\lq{}exotic" virtual particles with negative energy at the submicroscopic level down to the Planck scale that acts as the very substratum from which quantum black holes emerge to briefly become wormholes. This makes the background space to be quantized. With this set-up, the probability of finding a quantum particle in a particular state can be defined by $N$ number of minimum length (or $\chi$ number of fundamental energy), and given by the quantity $|\psi|^{2}$. This is similar to the conventional quantum formalism. However, the quantities used in TMI have such properties similar to the electromagnetic potentials like in Aharonov-Bohm Effect that is proven to physically exist. Also, the time remains a background parameter when used in the  description of fluctuating space via the Hyperbolic Ricci Flow. Lastly, the notion that complex number has physical meaning is needed just as the complex conformal (Weyl) transformation is needed as a fundamental symmetry in Nature.\\
\indent As a final note, in TMI, there is a way to intepret and visualize, in a physical way, why a quantum particle is behaving in a particular way. There is no defeatist mentality that the quantum world can never be explained in any physical and logical way as Copenhagen Interpretation suggests. There is indeed a physically existing wave (an oscillating space) such that the non-classical behaviour of a quantum particle can only be traced in the dynamics and nature of the surrounding space. \\

\clearpage

      \clearpage                                           
 % % % % % % % % % % % % % % % % % % % % % %
\section*{Conclusion}%Conclusion   
\indent An entropic and conformal gravity approach on the nature of Quantum Mechanics was presented based on the idea that at the Planck Scale, the spacetime is inherently quantized and fluctuating, where the Laws of Quantum Mechanics and General Relativity are modified. The quantized nature of the spacetime is described via an inclusion of a minimum length and energy scale. The fluctuation of spacetime at the Planck scale was described by a second-order version of the Ricci Flow with the metric tensor subjected to a complex Weyl transformation. The model offers a new approach to the resolution of the foundational problems of Quantum Mechanics with a new quantum formalism and interpretation based on the idea of Quantum Mechanics being emergent from the \lq{}\lq{}Physics at the Planck scale". The new quantum formalism suggests that everything that can be known in a quantum system is encapsulated not just in the wave function $\psi$ but also in a conformal metric tensor, $\bar{g}_{\mu\nu}= \psi^{2} g_{\mu\nu}$. The dynamics of a quantum system are shown to be related to the spacetime fluctuation that is co-varying with the property of a quantum field. In particular, it is dictated by the presence of submicroscopic wormholes that are stable in a short period of time, generated via the constant interaction with the surrounding empty space as a source of negative energy. Consequently, the probabilistic nature and \lq{}\lq{}non-classical" properties of Quantum Mechanics can be explained using some of the theorems and conjectures at the Planck scale, like the modified form of Area Theorem via the AdS/CFT conjecture. Quantum entanglement was shown to be an emergent property as prescribed by the ER=EPR conjecture. Similarly, the formation of the interference pattern in the double slit experiment was shown to be an emergent phenomena that can be attributed to the following factors: The passage of quantum particle through a series of wormholes where the number of exit points (i.e., wormhole mouths) increases in time, the probabilistic nature of where the exit points will appear, the chaotic effect of the particle's interaction with the barrier or the slits and the NP-hard combinatorial optimization to find the shortest possible path as dictated by the Principle of Least Action.
% % % % % % % % % % % % % % % % %
\section*{Acknowledgments}
\indent I am very much grateful to the NEU Board of Trustees and Administration, the NEU University Research Center and the NEU Press and Publication Office for all the support for the publication of this paper. 
%\section*{Bibliography}  
\fancyhf{}
%\fancyhead[R]{\textbf{NEW ERA SCIENCE \\ Vol. 1. Issue No. 1, June, 2025}}
\newpage
\fancyhf{}
\fancyhead[R]{\textbf{NEW ERA SCIENCE \\ Vol. 1. Issue No. 1, June, 2025}}
\bibliographystyle{spp-bst}

\bibliography{bibtopo1}

\end{document}